\documentclass[useAMS,usenatbib]{mn2e}

\usepackage{xspace}
\usepackage{hyperref}
\usepackage{aas_macros}
\usepackage{url}
\usepackage{amsmath}
\usepackage{amssymb}
\usepackage[capitalise]{cleveref}
\usepackage{color,graphicx}

\newcommand{\eobs}{\epsilon^{\mathrm{obs}}}
\newcommand{\eint}{\epsilon^{\mathrm{int}}}
\newcommand{\sersic}{S\'{e}rsic\xspace}
\newcommand{\imshape}{{\sevensize{IM3SHAPE}}\xspace}
\newcommand{\healpix}{{\sevensize{HEALPix}}\xspace}
\newcommand{\scubed}{{{S3-SEX}}\xspace}
\newcommand{\Clobs}{\hat{C}_\ell}

\newcommand{\nn}{\nonumber}

\newcommand{\om}{\Omega_\mr m}
\newcommand{\omb}{\Omega_\mr b}
\newcommand{\sig}{\sigma_8}

\newcommand{\w}{w_0}
\newcommand{\wa}{w_a}

\newcommand{\pd}{P_{\delta}}

\newcommand{\mr}{\mathrm}

\newcommand{\fsky}{f_{\rm sky}}
\newcommand{\E}{\times10}
\newcommand{\jbca}{{Jodrell Bank Centre for Astrophysics, School of Physics \& Astronomy, The University of Manchester, Manchester M13 9PL, UK}}

\title[SKA Weak Lensing II: Simulations]{SKA Weak Lensing II: Simulated Performance and Survey Design Considerations}
\author[Bonaldi et al.]{Anna Bonaldi\textsuperscript{\thanks{E-mail: a.bonaldi@skatelescope.org}}, Ian Harrison\textsuperscript{\thanks{E-mail: ian.harrison-2@manchester.ac.uk}}, Stefano Camera \& Michael L. Brown\\ \jbca}

\begin{document}

\pagerange{\pageref{firstpage}--\pageref{lastpage}} \pubyear{2015}

\maketitle

\label{firstpage}

\begin{abstract}
We construct a pipeline for simulating weak lensing cosmology surveys with the Square Kilometre Array (SKA), taking as inputs telescope sensitivity curves; correlated source flux, size and redshift distributions; a simple ionospheric model; source redshift and ellipticity measurement errors. We then use this simulation pipeline to optimise a 2-year weak lensing survey performed with the first deployment of the SKA (SKA1). Our assessments are based on the total signal-to-noise of the recovered shear power spectra, a metric that we find to correlate very well with a standard dark energy figure of merit. We first consider the choice of frequency band, trading off increases  in  number counts at lower frequencies against poorer resolution; our analysis strongly prefers the higher frequency Band 2 (950--1760\,MHz) channel of the SKA-MID telescope to the lower frequency Band 1 (350--1050\,MHz). Best results would be obtained by allowing the centre of Band 2 to shift towards lower frequency, around 1.1\,GHz. We then move on to consider survey size, finding that an area of 5,000 square degrees is optimal for most SKA1 instrumental configurations. Finally, we forecast the performance of a weak lensing survey with the second deployment of the SKA. The increased survey size (3$\pi$\,steradian) and sensitivity improves both the signal-to-noise and the dark energy metrics by two orders of magnitude.  

\end{abstract}

\section{Introduction}
Weak lensing analyses measure coherent distortions in the shapes of large numbers of background galaxies caused by the gravitational lensing effect of intervening matter along the line of sight. This enables both direct mapping of mass within the Universe (without recourse to assumptions about how luminous matter traces total matter) and tracking of the dark energy equation of state to enable constraints on physical models of the observed accelerated expansion. This science goal can be pursued through statistical analyses of the amount of lensing caused by large scale structure (or ``cosmic shear") and how it evolves with redshift, and is widely recognised as one of the most important outstanding questions in contemporary cosmology \citep[e.g.][]{2013PhR...530...87W}.

To date, the overwhelming majority of cosmic shear analyses have been performed in the optical waveband, from the first detections \citep{2000MNRAS.318..625B, 2000Natur.405..143W,2000A&A...358...30V,2000astro.ph..3338K} to the current state of the art represented by the CFHTLens \citep{2013MNRAS.432.2433H}, DLS \citep{2015arXiv151003962J} and DES-SV \citep{PhysRevD.94.022001} surveys. However, the effects of gravitational lensing are achromatic and a promising complementary pursuit is to measure weak lensing in the radio band, as already demonstrated over a decade ago by \cite{chang2004}. Within the next decade, the Square Kilometre Array\footnote{\url{http://www.skatelescope.org}}(SKA) will revolutionise the field of radio weak lensing as it will be the first radio telescope capable of detecting the large number densities of resolved, high redshift sources necessary for performing precision cosmology experiments.

Future weak lensing surveys with DES\footnote{\url{http://www.darkenergysurvey.org}}, KiDS\footnote{\url{http://kids.strw.leidenuniv.nl}}, HSC\footnote{\url{http://subarutelescope.org/Projects/HSC}}, SKA, LSST\footnote{\url{http://www.lsst.org}}, the WFIRST-AFTA\footnote{\url{http://wfirst.gsfc.nasa.gov}} and \emph{Euclid}\footnote{\url{http://euclid-ec.org}} satellites will be large and deep enough that it is knowledge of systematic effects (both instrumental and physical) on the measured cosmic shear, not statistical uncertainties, which will limit the precision of cosmological parameter constraints and subsequent confidence in model selection. Measuring the weak lensing signal from radio-wavelength data carries a number of potential unique advantages over other wavebands, as discussed in \cref{sec:radio_lensing} and \cite{2015arXiv150103828B}, which can be used to gain additional information and mitigate many of these systematics. Also, cross-correlations between shear maps created at different wavelengths can remove systematics, such as un-modelled beam ellipticity, which are not common between experiments \citep[as in][]{2015arXiv150705977D}.

In a companion paper \citep[][hereafter Paper I]{harrisoncamera2015}, we use Monte Carlo Markov Chain (MCMC) forecasting techniques to derive theoretical predictions for the scientific reach of SKA weak lensing surveys in terms of the forecasted constraints on cosmological models including dark energy and Modified Gravity theories. Paper I also examines in detail the statistical precision with which optical-radio cross-correlation techniques -- which should be very robust to instrumental systematic effects -- can constrain cosmological models (e.g. by combining the SKA2 and a Dark Energy Task Force Stage IV optical weak lensing survey like \emph{Euclid}).

In this paper we construct a detailed simulation pipeline in order to perform a more realistic assessment of the prospects for SKA weak lensing. While Paper I { presents the science case for radio weak lensing} and for optical-radio cross-correlation studies, the analysis we present here goes significantly beyond this and takes into account a number of real world effects that are not accounted for in the forecasts of Paper I---and indeed, which are also absent from most other weak lensing forecasting studies.

Our simulation pipeline takes as input cosmological shear power spectra and simulated radio source populations. We then create mock catalogues of objects according to selection cuts specified by the depth and resolution available from the envisaged SKA surveys at a specified observing frequency. We assign to each object a measured shape, taking into account a realisation of the cosmological shear signal (based on the supplied theory spectra) as well as additional contributions to mimic the effects of the intrinsic distribution of galaxy ellipticities and errors from the measurement of shapes in real data. In addition we assign each object an ``observed" redshift composed of the true redshift plus an error term. Our model for the redshift errors is based on the assumption that redshift information for the SKA source catalogue will be provided mostly by overlapping optical photometric redshift surveys. Whilst ultimately the full SKA will provide its own spectroscopic redshifts for many weak lensing sources through 21cm HI line surveys \citep[see e.g.][]{2015MNRAS.450.2251Y}, this will not be possible for the majority of the high-redshift lensing sources during the first phase of the SKA. The simulated object catalogues are subsequently passed through an analysis pipeline, composed of a multiple redshift-bin map-making step followed by a tomographic auto- and cross-power spectrum analysis. This latter step includes a correction for the limited sky coverage of the envisaged SKA surveys plus a noise bias subtraction step to correct the power spectra for the effects of galaxy ellipticity errors due to both the intrinsic scatter in galaxy shapes and measurement noise. 

As an example application of our simulation tool, we use our pipeline to determine the optimum observation frequency for an SKA weak lensing survey with SKA1-MID, the first phase of deployment of the mid-frequency dish interferometer component of SKA. Whilst source number counts are higher at lower frequency, due to the $\nu^{-0.7}$ synchrotron spectrum expected for radio emission from star-forming galaxies, the instrumental resolution decreases and point spread function (PSF) distortions from the turbulent ionosphere become more important. By exploring the parameter space in central observing frequency, PSF size and image-plane root-mean-square (RMS) noise level, and by defining a suitable figure of merit (FoM) that quantifies the overall precision of the power spectrum recovery, we are able to determine the instrument configuration which provides the best weak lensing performance. For the identified optimum configuration, we also investigate the optimum survey area for a fixed amount of SKA1-MID observing time. Finally, looking ahead, we also present a simulation of the weak lensing performance of the second phase deployment of the SKA (SKA2-MID) which will provide { a tenfold increase in survey depth \citep{2015aska.confE.174B}}, allowing for significantly improved cosmological constraints.

The paper is organised as follows. \Cref{sec:weak_lensing} provides an overview of weak lensing cosmology, with a particular focus on the advantages of radio weak lensing in \cref{sec:radio_lensing}. Survey and telescope specifications are laid out in \cref{sec:specs}. In \cref{sec:pipeline} we describe the SKA observation simulation pipeline and its ingredients. The application of our pipeline is demonstrated in \cref{sec:nu_results} where we also investigate the optimal instrumental configuration and survey areas for weak lensing with SKA1-MID.  
We present our conclusions in \cref{sec:conclusions}.
\section{Weak Lensing Cosmology}
\label{sec:weak_lensing}
Gravitational lensing is the distortion of the images of background sources by the deflection of light caused by the gravitational potential of massive objects along the line of sight \citep[see e.g.][and references therein for a full description]{schneiderbook}. In the weak lensing limit, the distortion of images can be written as a Jacobian transformation between the co-ordinate systems in the source and image plane, where the Jacobian matrix is 
\begin{equation}
A = 
\begin{pmatrix}
1 - \kappa - \gamma_1 & -\gamma_2 \\
-\gamma_2 & 1 - \kappa + \gamma_1
\end{pmatrix},
\end{equation}
where $\kappa$ is the convergence and $\gamma$ the complex shear, $\gamma = \gamma_1 + i\gamma_2$. The convergence represents the change in source size (whilst preserving surface brightness) and the shear is a spin-2 field which stretches circular light profiles into ellipses in one of two polarisations: one in parallel to a pair of chosen reference axes and one at 45 degrees to this. In the weak lensing regime the shear may be related to the integrated gravitational potential $\psi$ along the line of sight between the source and observer at sky position $\theta$:
\begin{align}
\gamma_1 &= \frac{1}{2}\left(\frac{\partial^2 \psi}{\partial\theta^2_1} - \frac{\partial^2 \psi}{2\partial\theta^2_2}\right), \nonumber \\
\gamma_2 &= \frac{1}{2}\frac{\partial^2 \psi}{\partial\theta_1 \partial\theta_2}.
\end{align}
Contributions to this potential come from all gravitating matter along the line of sight, meaning the cosmic shear is a probe of the full matter distribution directly, without recourse to bias prescriptions between visible and dark matter.

Elliptical light profiles of background sources (galaxies here) may also be mapped onto a spin-2 field, the ellipticity $\epsilon$. For simple galaxy profiles with elliptical isophotes, we have
\begin{align}
|\epsilon| &= \frac{1 - \left(b/a\right)^2}{1 + \left(b/a\right)^2}, \nn \\
\epsilon_1 &= |\epsilon| \cos{2\phi} \nonumber, \\
\epsilon_2 &= |\epsilon| \sin{2\phi},
\end{align}
where $a$ and $b$ are the galaxy major and minor axis and $\phi$ is the galaxy position angle (more generally, $\epsilon$ can be written as a function of second order moments in the galaxy light profile). When observed, a galaxy's ellipticity then receives contributions from the intrinsic (projected) shape of the source and the additional ellipticity generated by gravitational lensing shearing. In the weak lensing limit, we can model this as 
\begin{equation}
\eobs = \gamma + \eint,
\end{equation}
where $\eint$ is the galaxy's intrinsic ellipticity. Under the assumptions that, within a chosen cell on the sky, { the shapes of many galaxies are distorted by the same gravitational potential} and have intrinsic shapes which are uncorrelated with an azimuthally symmetric probability distribution (i.e. giving $\langle \eint \rangle = 0$) then the observed ellipticity averaged over $N$ galaxies within this cell becomes an unbiased estimator of the cosmic shear.
Even tighter constrains can be obtained by sub-dividing the samples into multiple tomographic redshift bins (labelled $i,j$) and considering the two-point statistic in spherical harmonic space:
\begin{equation}
C ^{ij} _\ell = \frac{9H_0^4 \om^2}{4c^4} \int_0^{\chi_\mr h} 
\mr d \chi \, \frac{g^{i}(\chi) g^{j}(\chi)}{a^2(\chi)} \pd \left(\frac{\ell}{f_K(\chi)},\chi \right).
\label{eq:cl_theory}
\end{equation}
Here, $H_0$ is the Hubble constant, $\om$
is the (total) matter density and $c$ is the speed of light. $a(\chi)$
is the scale factor of the Universe at comoving distance $\chi$ and
$f_K(\chi)$ is the angular diameter distance (given simply by
$f_K(\chi) = \chi$ in a flat Universe). $\pd(k, \chi)$ is the
matter power spectrum. The functions $g^{i/j}(\chi)$ are the lensing
kernels for the two redshift bins in question, which depends on the distributions of galaxies.

Using \cref{eq:cl_theory}, the
matter power spectrum for a given cosmology $\pd(k, \chi)$ may then be
related to the observed shear power spectrum,
\begin{equation}
\Clobs = \frac{1}{2\ell + 1} \sum_{m=-\ell}^{\ell} a^i_{\ell m} a^{j*}_{\ell m}
\label{eq:clobs}
\end{equation}
where $a^{i}_{\ell m}$ are the $E$-mode coefficients in a spin-2 spherical harmonic
expansion of the observed shear field for bin $i$, and the asterisk denotes complex conjugation. 

To first approximation, the achievable errors on $\Clobs$ scale simply
with the survey area $\fsky$, the  sky number density of galaxies available $n_{\rm gal}$ and
the RMS variance of the galaxies' ellipticity distribution
$\sigma_{\epsilon}$:
\begin{equation}
\label{eqn:delta_cl}
\Delta \Clobs = \sqrt{\frac{2}{(2\ell + 1)\fsky}}\left(\Clobs + \frac{\sigma_{\epsilon}^2}{n_{\rm gal}} \right).
\end{equation}
Given a set of observed shear power spectra and an associated
covariance matrix, one can then use the above formalism to compare
weak lensing observations to the expected signal for a given cosmology
and estimate the most likely values of parameters within that
cosmological model.

\subsection{Radio weak lensing}
\label{sec:radio_lensing}
Up to now, weak lensing measurements at radio wavelengths have been limited by surveys' ability to achieve the high number densities of high redshift background sources necessary. One exception is the study using data from the Faint Images of the Radio Sky at Twenty centimetres (FIRST) survey by \cite{chang2004}: a wide, shallow survey in which a 3.6\,$\sigma$ detection of a lensing signal was made. \cite{patel2010} also measured the shapes of objects, in combined Very Large Array (VLA) and Multi-Element Radio Linked Interferometer Network (MERLIN) observations of a $70 \, \mathrm{arcmin}^2$ region of the Hubble Deep Field-North, in both radio and optical, at high number densities but with an overall sample size too small for a weak lensing detection. Most recently, \cite{2015arXiv150705977D} cross-correlated galaxy shapes in the radio FIRST survey with those in the optical Sloan Digital Sky Survey (SDSS) and made a marginal detection of cosmic shear, whilst demonstrating the removal of wavelength-dependant systematics.

Newly upgraded radio telescopes such as the JVLA and e-MERLIN are capable of the sensitive, wide field-of-view observations necessary for a weak lensing survey and the forthcoming SKA will provide cosmological constraints competitive with premier optical experiments \citep{2015arXiv150103828B, harrisoncamera2015}. The Super-CLuster Assisted Shear Survey (SuperCLASS\footnote{\url{http://www.e-merlin.ac.uk/legacy/projects/superclass.html}}) is an e-MERLIN legacy survey whose primary science goal is to measure a significant weak lensing signal in the radio, learning about the properties of relevant background sources and developing techniques for shear measurement in interferometric data.

In addition to simply providing competitive background source number densities and survey areas, radio weak lensing surveys have a number of unique advantages \citep{2015arXiv150103828B}. { Firstly, radio interferometer PSFs are in principle highly deterministic and stable in time, removing potential biases in source shape reconstruction.
Moreover, an SKA survey will access larger angular scales (because of the large fraction of the sky covered, up to $\fsky = 0.75$) and higher redshifts (because of the source redshift distribution in the radio) than achievable in other wavebands, thus providing better constraint on dark energy models and least dependence on poorly-known non-linear scales.
Finally, it has been argued that polarisation \citep{brown2011, 2015MNRAS.451..383W} and rotational velocity \citep{blain2002, morales2006, huff2013} information available in the radio could be used to trace the intrinsic alignment of objects \citep[see e.g.][]{kirk2015}. }

The extent to which these potential advantages will be realised depends on both the exact properties in continuum and polarisation of the $\mu\mathrm{Jy}$ radio source population and the development of necessary data analysis techniques. Nevertheless, information uniquely available from radio surveys will inevitably add more constraining power to weak lensing cosmology and provide a valuable cross-check.
In particular, cross-correlations between shear maps constructed from radio and optical surveys can remove systematic errors induced in each map by the telescope, as explored in \cite{camera2016}.

\section{Telescope and Survey Specifications}
\label{sec:specs}
In this section, we summarise the instrumental specifications of the SKA that are relevant for weak lensing. We use these specifications to model the instrument in the simulation pipeline described below. The SKA \citep[see chapters within][]{2015aska.confE.174B} will be built between now and the late 2020s, with the current plan consisting of two main phases of construction, both taking science quality data. Here we focus on the first phase of the SKA (SKA1) expected to be complete in 2023, and with early science observations (with approximately 50\% of the array) beginning around 2020. SKA1 will consist of two sub-arrays: SKA1-LOW will be an aperture array operating at low radio frequencies from $50\,$MHz to $350\,$MHz and built in Western Australia, whilst SKA1-MID will be a dish array with up to five observational frequency bands spanning the range $350\,$MHz to $13.8$\,GHz located in Southern Africa. For weak lensing, it is SKA1-MID which is of interest, as it provides both the sensitivity and spatial resolution to capture morphological information on high redshift ($z\gtrsim1$ ) star forming galaxies which are used for measuring cosmic shear. SKA1-MID will represent a significant jump in the capabilities of radio telescopes, improving on the JVLA by factors of 4 in resolution, 5 in sensitivity and 60 in survey speed.

The observing frequency range for SKA1-MID will be split into multiple frequency bands. As currently defined, the bands of interest for radio weak lensing are Band 1 and Band 2. Band 1 will cover $350\,$MHz to $1050\,$MHz (giving a bandwidth of $700\,$MHz) and Band 2 will cover $950\,$MHz to $1760\,$MHz (giving a bandwidth of $810\,$MHz). These band definitions are subject to change and, in this work, one of the questions we attempt to answer is what the optimal central frequency would be (for a fixed overall bandwidth) in order to maximise the weak lensing science return. To address this, we will consider central observing frequencies across the whole range $350\,$MHz to $1760\,$MHz (i.e. we do not require that the entire frequency coverage considered within a simulated experiment sits solely within one band). 

As a default we consider a usable bandwidth of 30\% of the total available bandwidth in the current definition of Band 2. We note that this is a conservative approximation: in principle, one could use the entire $810\,$MHz bandwidth in order to extract galaxy shape measurements since the lensing signal we are interested in is achromatic. Moreover, the colour gradient bias caused by un-modelled frequency-dependent PSF effects \citep{2012MNRAS.421.1385V} is expected to be much less of a problem in radio observations because frequency-dependent PSF effects can be modelled exactly. However, even though the SKA telescope will be situated in an exceptional Radio Frequency Interference (RFI) quiet site, in practice parts of the band may still be rendered unusable due to RFI caused by mobile phones, satellites etc. 

In all of our simulations, we adopt a total amount of observing time on SKA1-MID of 10,000 hours. In addition to weak lensing, such a survey would meet many other continuum and cosmological science goals and could be done commensally with a 21cm HI line survey. We also note that the sensitivity of the observations scales with observing time and bandwidth in an identical fashion. Thus, the conclusions we draw from our adopted 30\% bandwidth / 10,000 hr configuration are also immediately applicable to a 100\% bandwidth / 3,000 hr configuration. 

In order to explore the optimal central frequency for weak lensing, we consider a default survey area of 5,000 square degrees. For a limited number of well performing central frequency choices, we also investigate the dependence of the weak lensing science performance on the survey area. 

\section{Simulation Pipeline}\label{sec:pipeline}
The simulation pipeline we constructed consists of the following steps:
\begin{enumerate}
\item Input shear power spectra are computed according to a given cosmology and for a set of redshift slices.
\item Simulated shear maps are generated based on the input power spectra.
\item A catalogue of galaxies is generated depending on intrinsic properties (joint flux, redshift and size distribution for the considered galaxy populations) and instrumental specifications (frequency, resolution, sensitivity, sky coverage of the observation).
\item A measured redshift is associated to each galaxy of the catalogue, with different measurement errors adopted depending on whether a photometric or spectroscopic redshift is being modelled.
\item A measured ellipticity is associated to each galaxy in the catalogue, with shear ellipticity given by the simulated shear map at the redshift and sky position of the galaxy, plus intrinsic ellipticity and shape measurement errors.
\item The simulated galaxy ellipticity catalogues are binned into shear maps from which the shear power spectra are estimated.  
\item The recovered power spectra are averaged over several shear and catalogue realisations and are compared with the equivalent expected power spectra from theory.
\end{enumerate}
Each step of the pipeline is described in detail in the following sub-sections. 

\subsection{Input shear power spectra and maps}
\label{sec:input_shear_model}
For this paper we consider a set of best-fitting cosmological parameter values from the Planck 2015 results. Specifically, we adopt the  ``TT,TE,EE+lowP+lensing+ext" results from \cite{2015arXiv150201589P}: $\lbrace  \om, \sig, h, \omb, \w, \wa \rbrace = \lbrace 0.309, 0.816, 0.677, 0.0487, -1, 0 \rbrace$. Here, $\sigma_8$ is the normalisation of the matter power spectrum expressed in terms of the RMS density fluctuations averaged in spheres of radius 8 $h^{-1}$Mpc and $h$ is the dimensionless Hubble parameter defined such that $H_0 = 100\,h$ km~s$^{-1}$~Mpc$^{-1}$. $\om$ and $\omb$ are the matter and baryon density parameters while $w_0$ and $w_a$ are the equation of state of dark energy and its redshift dependence. 

In this paper, we choose to model the 3-dimensional (3D) cosmological shear field as a series of correlated 2-dimensional (2D) shear fields at multiple positions in redshift space. This approach allows us to efficiently generate realisations of appropriately correlated Gaussian random shear fields, on the full (curved) sky, at multiple positions in redshift space. We choose the number and redshift positioning of our 2D shear fields to match the median redshifts of a set of narrow redshift bins, which are in turn chosen to ensure that the size-flux distribution of our modelled galaxy population is approximately constant across each bin (see \cref{sec:source_pop_model}). This prescription defines a set of $n_{\rm bin} = 11$ median redshifts spanning the range $0.036 < z_m < 10$ at which we generate the input shear fields.

The auto- and cross- shear power spectra for each redshift bin combination are calculated according to \cref{eq:cl_theory}. Our power-spectrum generation code uses the \cite{1984ApJ...285L..45B} approximation to calculate the linear CDM power spectrum, $P_\delta(k, \chi)$, and we use the {\sevensize HALOFIT} code \citep{2003MNRAS.341.1311S} to calculate the non-linear $P_\delta(k, \chi)$.  

Our simulation proceeds by generating a set of $n_{\rm bin}$ Gaussian random shear fields, properly correlated between different bins (according to the  appropriate shear cross-spectra) following the procedure described in \cite{brown2011}. Briefly, at each multipole $\ell$, we form the $n_{\rm bin} \times n_{\rm bin}$ power spectrum matrix, $C_\ell^{ij}$. Taking the Cholesky decomposition of this matrix ($L_\ell^{ij}$) defined by  
\begin{equation}
C_\ell^{ij} = \sum_z L_\ell^{ik}L_\ell^{jk}, 
\end{equation}
we generate random realisations of the spin-2 spherical harmonic coefficients of the shear fields on each redshift slice according to
\begin{align}
a^i_{\ell 0} &= \sum_y L^{ij}_\ell G^j_{\ell 0}, \nn \\
a^i_{\ell m} &= \sqrt{\frac{1}{2}} \sum_y L^{ij}_\ell G^j_{\ell m},
\label{eq:alm_sim_fields}
\end{align}
where $G^i_{\ell m}$ is an array of unit-norm complex Gaussian random
deviates. We then use the \healpix software \citep{gorski05} to transform the shear field for each redshift slice to real space maps of $\gamma_1$ and $\gamma_2$. Note that the harmonic modes of \cref{eq:alm_sim_fields} are the even-parity $E$-modes. In this work we model the lensing shear fields as pure $E$-mode, and consequently, the $B$-modes are set to zero. Our numerical implementation employs a \healpix resolution parameter $N_{\rm side} = 2048$ which corresponds to a shear map pixel size of $\sim2.7$ arcmin. Our subsequent power spectrum recovery will therefore be limited to $l_{\rm max} \leq 2N_{\rm side} = 4096$. 

\subsection{Source populations}
\label{sec:source_pop_model}
Below $\sim$1\,mJy the 1.4\,GHz source counts change from being dominated by active galactic nuclei (AGNs) to being dominated by star-forming galaxies \citep[see, e.g.][]{hopkins03,muxlow05,moss07,condon12}, even though radio-quiet AGNs could still contribute to the upturn below 1\,mJy \citep{jarvis04,white15}. In our simulation we have attempted to mimic a weak lensing analysis that makes use of the star-forming galaxy population only.

To model the properties (flux, size and redshift distributions) of the star-forming population we relied on the SKA design studies S-cubed Extragalactic (\scubed) simulation of \cite{wilman08}, after comparing the simulation outputs with the most recent data and performing a few adjustments. 

In particular, at the time the \scubed simulation was produced there was very little information on the source counts below the mJy level, where star-forming galaxies begin to dominate. The situation has significantly improved over the last years, thanks to deeper surveys and the use of statistical methods to constrain source counts below the detection threshold, such as stacking \citep{dunne2009, karim2011}, maximum likelihood \citep{mitchell2014} and ``$P(D)$'' \citep{condon12, vernstrom14}. 

\begin{figure}
\begin{center}
\includegraphics[width=0.5\textwidth]{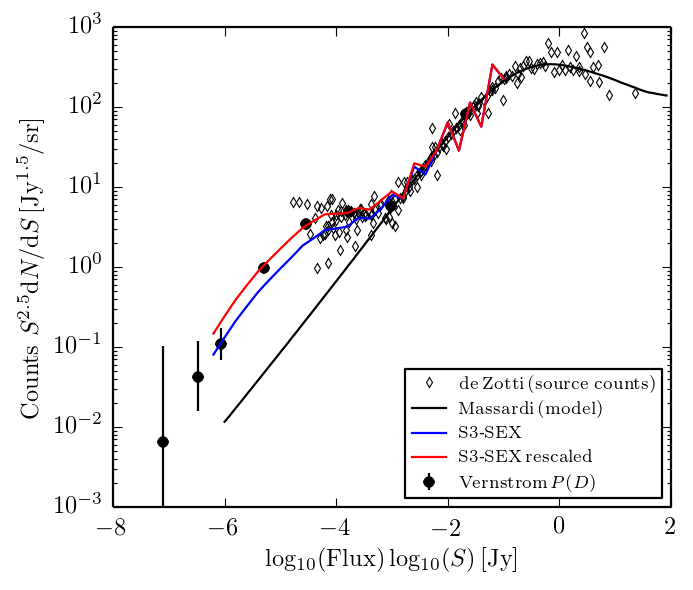}
\end{center}
\caption{Source counts at 1.4 GHz: comparison between the compilation of source counts data in \citet{2010A&ARv..18....1D} (diamonds), the ``$P(D)$'' analysis of \citet{vernstrom14} (points with error bars), the simulation from \citet{wilman08} (blue solid line) and its rescaled version (red line). The error bars of the $P(D)$ analysis are smaller than the plotting symbol above $10^{-6}\, \mathrm{Jy}$. For this work, we re-scale the star-forming galaxies with a factor of $2.25$ which, as shown in the figure, provides a much better fit to the observations.}
\label{fig:counts_motivation}
\end{figure}

Since this is clearly a key ingredient of the simulation, we recalibrated the number of star-forming galaxies in \scubed, by means of a global normalisation factor, to match the most recent deep radio observations.
To obtain this factor we computed the differential source counts of the \scubed star-forming galaxies and compared them with the $P(D)$ analysis of \citet{vernstrom14} between 5 and 200\,$\mu$Jy, corrected for the AGN contribution with the best-fit model of \citet{massardi10}. 
Based on this analysis we multiply the number of galaxies sampled from the distributions in \scubed by a factor of 2.25, as we believe this best represents the true number densities available to SKA1-MID (see \cref{fig:counts_motivation}). { This recalibration is fully consistent with other statistical analyses \citep[e.g.][]{condon12} and improves the agreement with several other deep radio studies that are based on number counts of galaxies detected at high $S/N$ \citep{muxlow05, 2010ApJS..188..178M, 2010ApJS..188..384S}. }

\begin{figure}
\begin{center}
\includegraphics[width=0.5\textwidth]{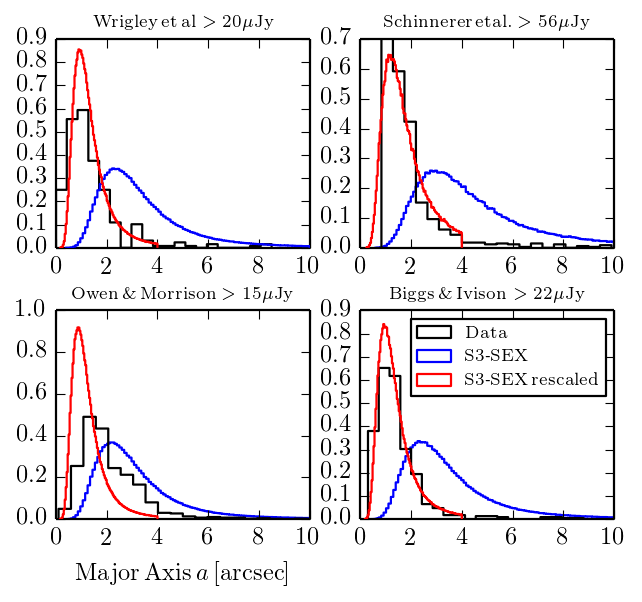}
\end{center}
\caption{Available angular size data of faint radio sources and comparison to the distribution within the SKADS \scubed simulation \citep{wilman08}. The sizes of objects in \scubed appear to be significantly overestimated. For this work, we re-scale all object sizes with a factor of $1/2.5$ which, as shown in the figure, provides a much better fit to the observations.}
\label{fig:size_motivation}
\end{figure}

Similarly, we have re-calibrated the size information of \scubed in light of recent data from a number of deep radio experiments \citep[][Wrigley et al. in prep]{2006MNRAS.371..963B, 2008AJ....136.1889O, 2010ApJS..188..384S}. In \cref{fig:size_motivation} we compare the major axis full width at half maximum (FWHM) size distributions within these observed data sets to the comparable distributions at the same frequency and flux limit within SKADS \scubed. As can be seen, \scubed consistently over-estimates the sizes of objects, with a division by a factor 2.5 of all source sizes providing significantly better agreement with the data.

\begin{figure*}
\begin{center}
\includegraphics[width=\textwidth]{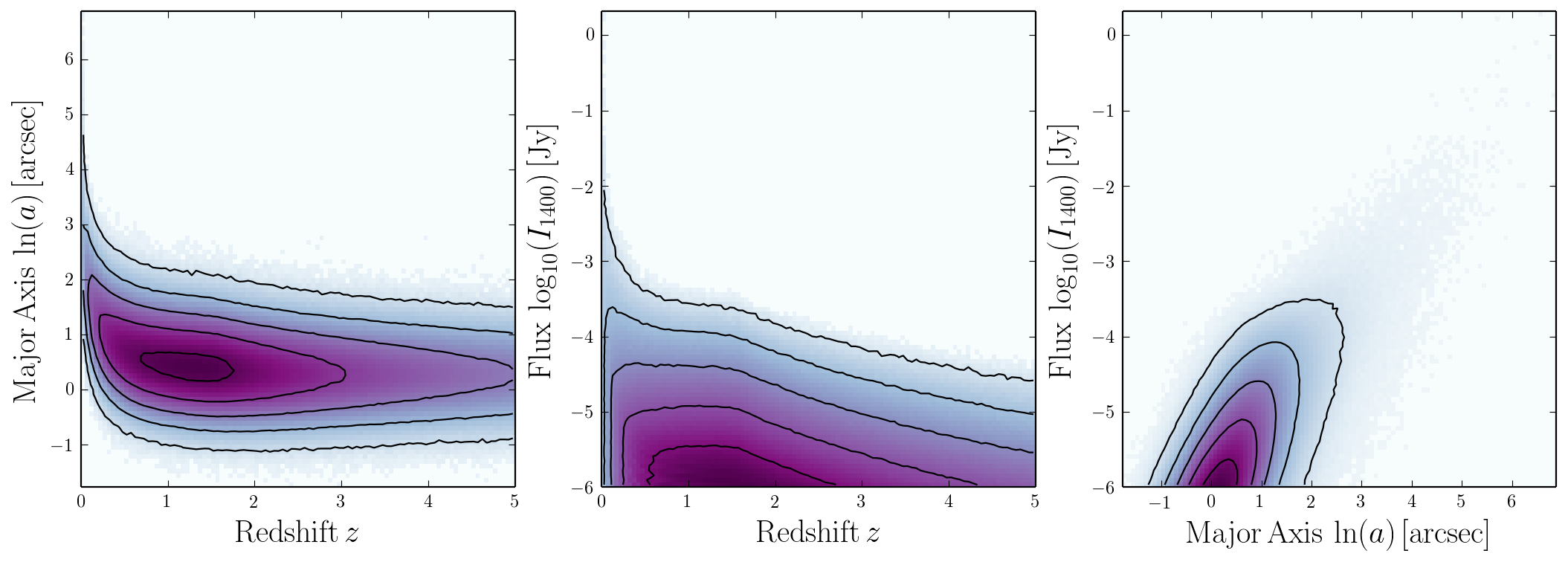}
\end{center}
\caption{Source angular size ($a$), redshift and 1.4\,GHz flux ($I_{1400}$) distributions taken from the SKADS S3-SEX simulation \citep{wilman08} and used for our simulation, with re-scalings as described in \cref{sec:source_pop_model}.}
\label{fig:3ddistr}
\end{figure*}

The projected 2D distributions in flux, size and redshift for star-forming galaxies within \scubed (after the above corrections have been applied) are displayed in \Cref{fig:3ddistr}. Given our procedure of simply rescaling quantities by a constant factor, the correlation between flux, size and redshift is the same as the original \scubed.  
In principle, a more consistent approach would be to update the simulation inputs (luminosity functions, redshift distributions, size models) and produce a new realization, rather than recalibrating its outputs. Such analysis is however beyond the scope of this work.

For generating our simulated catalogues, we should in principle sample from the joint 3D distribution of these three correlated quantities. 
However, for computational convenience we have defined narrow redshift bins to sample within which guarantee a weak dependence of the flux-size distribution on $z$ within a redshift slice. This allows us to simplify our simulation by sampling from the 2D flux-size distribution for each redshift bin independently. This procedure also sets the redshifts at which we simulate the correlated 2D shear fields, as previously described in \cref{sec:input_shear_model}. 

The generation of source samples from the 2D flux-size distributions is performed using the \scubed simulated fluxes at $\nu = 1.4\,$GHz.
The sampled sources are then propagated to the observational frequency of interest assuming a synchrotron scaling law $\nu^{-0.7}$. 

To sample from the 2D flux-size distributions, we first perform a principal component analysis on the flux-size samples from S3-SEX to get a set of two (maximally) uncorrelated new variables. We then sample from the 1D distributions of each of those two variables independently and finally convert these samples back to correlated flux and size with the inverse principal component transformation. The procedure to sample from a generic 1D distribution (in our case redshifts, principal components) is based on a change of variable to map the quantity of interest to a new one that is uniformly distributed. The mapping therefore depends on the probability distribution of the variable of interest. 
Once generated, a galaxy is included in the catalogue for a given experimental configuration according to two cuts:
\begin{itemize}
\item Flux cut: we require the total galaxy flux at the observational frequency to be greater than 10 times the RMS image-plane noise level ($S_{\rm rms}$) for the experiment.
\item Size cut: the galaxy major axis must be greater than 1.5 times the PSF FWHM for the experiment.
\end{itemize}
These cuts are conservative in comparison to those typically used for source catalogue creation, but are in line with the requirements for obtaining reliable galaxy ellipticity measurements for cosmic shear \citep[e.g.][]{2015arXiv150705603J}. 

In order to model the redshift estimates that we can expect to be available for future SKA surveys, we assume a fraction $f_{\textrm{spec-}z}=0.15$ of sources with redshift $z<0.6$ to have perfect redshift determinations from the spectroscopic detection of their HI line. For the remainder we assign a measurement error due to photometric redshift estimation which we assume can be supplied by overlapping wide-field optical surveys. 
Alternatively, further redshift information may be available for sources which do not meet the strict threshold for spectroscopic detection, via sub-threshold estimators which return a still useful posterior probability distribution for the redshift $P(z)$ \textit{\`a la} photometric estimators in the optical waveband.
For galaxies up to a maximum redshift of $z=2$, we draw from a Gaussian of width $\sigma_{\textrm{photo-}z} = 0.04(1+z)$. For galaxies at higher redshift, we draw from a broader Gaussian with $\sigma_{\textrm{no-}z}=0.3(1+z)$ { (see Paper I for the most conservative case where no redshift can be obtained for high-$z$ objects)}. 

\subsection{SKA survey sensitivities}
To implement the selection cuts on flux and size, for each experimental configuration (i.e. for each central observing frequency and survey area considered) we require an estimate of the effective sensitivity and resolution of the survey. This is straightforward to calculate for a direct imaging telescope as the sensitivity and resolution are uniquely determined by the instrumental design and observation frequency. For an interferometer telescope such as the SKA, the situation is slightly more complicated as these two quantities (the sensitivity and resolution of the telescope) are not independent. An interferometer is sensitive to the sky signal on a range of spatial scales which are determined by the array's distribution of baselines. The telescope's sensitivity to a particular spatial scale (or sky Fourier mode) is proportional to the number of corresponding baselines in the array while the size and shape of the effective PSF are also determined by the distribution of baselines and the particular array configuration. 

\begin{figure}
\begin{center}
\includegraphics[width=0.5\textwidth]{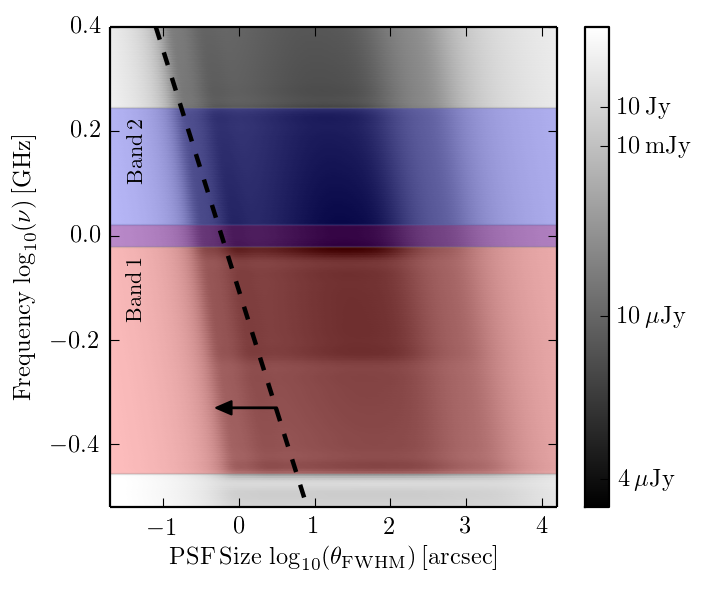}
\end{center}
\caption{The SKA1-MID sensitivity surface as a function of central observing frequency and image-plane PSF size. Regions to the left of the dashed line are not considered in this study, as the Kolmogorov PSF size is greater than the PSF size available from the instrument (see \cref{sec:ionosphere}).}
\label{fig:rb_sens_plot}
\end{figure}

For our purposes, the net effect of this is that for any given central observing frequency, there is an entire range of combinations of effective sensitivity and resolution that one can consider. To include these effects in our study, we use the most recent SKA1 Level 0 Science Requirements document produced by the SKA office \citep{skalevel02015}. For the frequency range of interest (essentially spanning the SKA1-MID Band 1 and 2 frequency ranges) that document presents estimates of the SKA1-MID sensitivity performance for a range of effective resolutions (or PSF sizes). The corresponding sensitivity surfaces are displayed, along with the overlayed frequency ranges of Band 1 and Band 2 (as currently defined) in \cref{fig:rb_sens_plot}.
These calculations assume a particular SKA1-MID antenna configuration (with maximum baseline length $150\,$km) and have been calculated for the case of a $10,000\,$hour wide-field survey making use of $30\%$ of the total bandwidth available within each SKA frequency band. The 30\% bandwidth assumption simplifies the interpretation for science areas that are sensitive to the spectral dependence of sources across the full band, and also accounts for RFI losses.\footnote{We note again that a weak lensing analysis could in principle extract useful information from the entire RFI-free bandwidth.} For each central frequency considered, the RMS image-plane noise level is then calculated in Jy beam$^{-1}$ for a survey covering $3\pi\,$steradians of the sky, with information on available baselines weighted to give a high-quality image-plane PSF of a given FWHM. In order to scale these sensitivities to different total sky areas for a fixed amount of observing time, we keep the quantity $\sqrt{A_{\rm sur}}/S_{\rm rms}$ fixed, where $A_{\rm sur}$ is the survey area and $S_{\rm rms}$ the RMS sensitivity. 

It may be expected that shape measurement for radio weak lensing will be conducted not in the image plane but directly with the raw Fourier-plane visibility data which constitutes the output from an interferometer \citep[see e.g.][]{2015arXiv150706639H}. Though it may not be optimal for weak lensing, the image-plane sensitivity at a given PSF size, displayed in \cref{fig:rb_sens_plot}, nevertheless provides a measure of the spatial scales on which the SKA1-MID telescope has baselines to measure morphological information, and we use it here in lieu of detailed simulations of visibility plane data.

\subsection{Population shape measurement}
For the galaxies meeting our flux and size selection cuts and being recorded in a catalogue for a given experiment, we also generate an observed ellipticity $\eobs$, given by the intrinsic ellipticipy plus a noise term. Noise on cosmic shear measurements consists of both shape noise (due to the intrinsic ellipticities of the source galaxies) and  measurement error. { The latter depends on the quality of the data (like signal-to-noise and resolution) but also on the method used to measure the shapes. Indeed, measuring ellipticities of individual sources within real data is a complicated image analysis task, whose development for radio interferometers is only just beginning \citep[see e.g.][and the radioGREAT challenge]{patel2014}.\footnote{\url{http://radiogreat.jb.man.ac.uk}} The status of the analysis in the optical is currently much better, thanks to the progress made between initial \citep{heymans2006} and recent \citep{mandelbaum2015} community algorithm challenges. 

Here, we assume that the quality of shape measurement that is currently state-of-the-art for optical weak lensing will be obtainable for future (2020 and later) SKA1 data. The requirements on systematic shear measurement errors in order for an SKA1 experiment to saturate its statistical error bars are also similar to the DES-like optical experiments whose measurement algorithm performance we consider here \citep[see Table 3. of][]{2015aska.confE..30P}.
To this end, we make use of large scale simulations developed for optical shear measurements \citep{Kacprzak} to model the total (noise plus measurement) shape errors.} These simulations take models of galaxy light profiles from deep optical observations of the COSMOS field by the Hubble Space Telescope. After PSF deconvolution the galaxies have their light profiles modelled with a shapelet decomposition \citep[as described in][]{2012MNRAS.420.1518M}. The galaxies are then placed in postage stamps with different noise realisations, and with amplitudes corresponding to a large range of signal-to-noise (SNR) values. \imshape, a shape measurement code which performs a maximum likelihood fit of \sersic profiles to a galaxy \citep{zuntz2013}, is then applied to these noisy postage stamps to simulate the effect of observing the ellipticities. These ``observed ellipticities'' contain both shape noise from the intrinsic galaxy ellipticity distribution already present in the \cite{2012MNRAS.420.1518M} catalogue as well as measurement noise from the \imshape fitting.

\begin{figure}
\begin{center}
\includegraphics[width=0.5\textwidth]{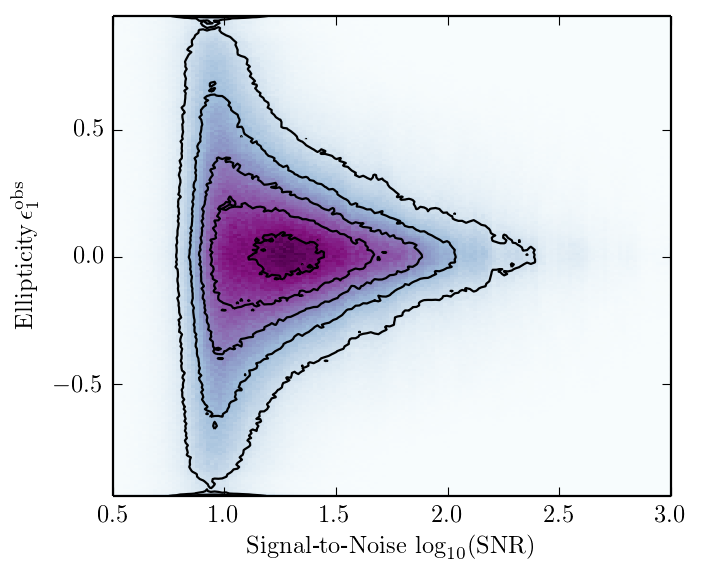}
\end{center}
\caption{Noise distribution for observed ellipticity, combining shape noise from intrinsic galaxy ellipticities and measurement noise from simulated \imshape measurements, showing the strong dependence on galaxy signal-to-noise ratio.}
\label{fig:tk_sims}
\end{figure}

\Cref{fig:tk_sims} displays a histogram of the ``observed ellipticities" ($\eobs$) generated for the  $\epsilon_{1}$ shear component (the $\epsilon_{2}$ component is similar), showing the strong dependence of the shear measurement error on the source SNR. We stress again that its validity for our purposes relies on the hypothesis that the quality of radio shape measurements for SKA1 will reach the level that is currently state-of-the-art in optical studies. 
In the case where this is achieved via shape measurement of individual galaxies in reconstructed images with approximately white noise properties, \imshape is a viable option and \Cref{fig:tk_sims} is directly applicable. Departures from this regime will modify the shape of the observed correlation between ellipticity and signal-to-noise. However, a strong trend between these two quantities, qualitatively similar to that of \Cref{fig:tk_sims}, has also been found in the extremely different regime of measuring radio band shapes directly from the visibilities \citep[Figure 5 of][]{2016MNRAS.463.1881R}.

Other source properties, such as angular size, are expected to scale the distribution, but more weakly and are ignored here. In our simulation, when a galaxy is drawn from the size-flux-redshift distributions, it is then assigned an observed ellipticity which is composed of the sum of the cosmological shear signal, $\gamma$ (generated as described in \cref{sec:input_shear_model}) and a sample drawn from the one-dimensional distribution extracted from \cref{fig:tk_sims} at the corresponding SNR. 
This ensures that the dependence of shape measurement uncertainties on source SNR is correctly accounted for in our simulations. 

\subsection{Ionospheric effects}
\label{sec:ionosphere}
A posited advantage of performing weak lensing analyses in the radio band is that, in contrast to ground-based optical experiments, turbulence on arcsecond scales in the Earth's atmosphere is not expected to significantly distort galaxy images and diminish the accuracy of shear measurements at radio frequencies. This `seeing' effect is a major limiting systematic for ground-based optical weak lensing experiments. However, at lower radio frequencies, turbulence in the ionosphere and troposphere means this seeing effect returns. The strong negative (typically $\sim \nu^{-2}$) scaling of these effects with frequency means they may be avoided by observing at higher frequencies. On the other hand, the $\nu^{-0.7}$ scaling of the dominant synchrotron emission from star-forming galaxies means that available galaxy number densities will also decrease at higher frequencies, thus increasing the shot noise term in \cref{eqn:delta_cl} and reducing the constraining power of cosmic shear measurements. One of the key uses of our simulation pipeline is to quantify the relative importance of these competing effects.

Whilst schemes exist for calibrating ionospheric effects \citep[see e.g.][]{2014arXiv1402.4889I}, they typically require interpolation of PSF corrections across the sky between the bright point sources which are used for calibration. Analogously in the optical waveband, PSF corrections are often interpolated between stars used as point sources \citep[e.g.][and the methods described therein]{2013ApJS..205...12K}. Unfortunately, the bright point sources required for ionospheric calibration do not appear at high densities on the sky, meaning residuals from corrections will likely be large compared to the shear signal used for weak lensing cosmology.

\begin{figure}
\begin{center}
\includegraphics[width=0.47\textwidth]{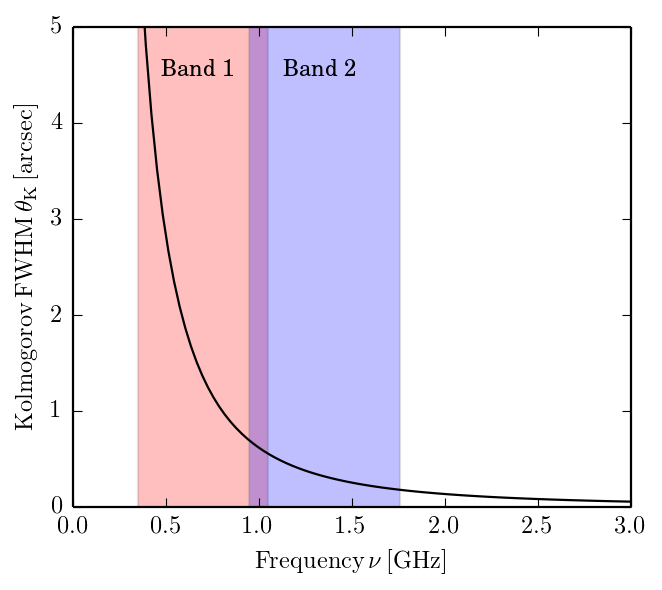}
\end{center}
\caption{The ionospheric PSF model used in this paper, which sets the limiting smallest achievable PSF for a given observational frequency. SKA1-MID Band 1 and Band 2 frequency regions are also marked.}
\label{fig:kolmogorov_psf}
\end{figure}

In order to include the effects of the ionosphere in our simulation pipeline, we use a simple model of Kolmogorov turbulence and take a conservative approach, expecting that no calibration is performed. 
For our Kolmogorov turbulence model we use a Fried parameter of $r_{d} = 10\,$km (measured at $150\,$MHz) from \cite{2015MNRAS.453..925V}. The FWHM of the Kolmogorov PSF is then given by:
\begin{equation}
\theta_{\mathrm{K}} \approx 0.976 \frac{\lambda}{r_{d}}
\end{equation}
with the Fried parameter scaling as:
\begin{equation}
r_{d} \propto \nu^{6/5}.
\end{equation}
The steep behaviour of the size of this PSF can be seen in \cref{fig:kolmogorov_psf}, along with the Band 1 and Band 2 frequency ranges. In regions of the frequency-PSF FWHM plane (\cref{fig:rb_sens_plot}) where the size of the Kolmogorov turbulence PSF, $\theta_K$, becomes greater than the available PSF FWHM from the experimental configuration, we discard the configuration. As shown in \cref{fig:rb_sens_plot}, this excludes a significant proportion of the parameter space covered by the currently envisaged SKA-MID Band 1 and a smaller proportion of the Band 2 parameter space.

\subsection{Shear maps and power spectrum estimation}\label{sec:shape}
For a given experimental configuration, the output of the above simulation pipeline consists of a simulated galaxy catalogue which contains for each object its angular co-ordinates on the sky, an estimated (noisy) redshift and estimated (noisy) ellipticity components. We then process these simulated catalogues through a tomographic shear power spectrum analysis. 

We first bin the ellipticity measurements in both angular co-ordinates and in redshift space. As already mentioned, the catalogue simulation pipeline was run for a set of 11 bins, spanning $0 < z < 10$, driven by computational requirements. For the power spectrum estimation, we rearranged them into 6 wider redshift bins, roughly equally-populated \citep[see, e.g.][for how to combine redshift bins]{Hu:1999ek}.
For the angular binning we again use the \healpix pixelization framework to construct estimated shear ($\gamma_1$ and $\gamma_2$) maps by simple averaging of the catalogued ellipticities within each sky pixel. To reduce the number of holes in the map for this step we use a \healpix resolution parameter $N_{\rm side} = 1024$ corresponding to a pixel size of $\sim5.4$ arcmin. 

The power spectrum estimation uses the \healpix package to transform the $\gamma_1$ and $\gamma_2$ maps to the spherical harmonic domain and to compute auto- and cross-spectra for all of the redshift bin combinations. These pseudo power spectra are calculated from the $E$-mode spherical harmonic coefficients according to \cref{eq:clobs}. The power spectra are then corrected for the effects of incomplete sky coverage and intrinsic shape and ellipticity measurement errors, and averaged in bandpowers (labelled $b$) according to
\begin{equation}
{\mathbf P}^{ij}_b = \sum_{b'} \left({\mathbf M}^{ij}_{bb'}\right)^{-1} \sum_\ell P_{b'\ell} (\hat{C}^{ij}_\ell - \langle N^{ij}_\ell \rangle_{\rm MC}).
\label{eq:cl_estimator}
\end{equation}
In \cref{eq:cl_estimator}, the matrix ${\mathbf M^{ij}_{bb'}}$ corrects observed power spectra for the effects of the finite survey area and the spatial distribution of simulated galaxies, while $\langle N^{ij}_\ell \rangle_{\rm MC}$ is the ``noise bias" due to intrinsic galaxy shape noise and measurement errors. We estimate this latter term from Monte Carlo simulations containing only noise. By randomising the position angle of the galaxies in the map, the shear signal is averaged out and we are left with just the effects of noise. Such noise power spectra are averaged over many (a total of fifty) realisations, until convergence is reached. Note that the noise bias must be subtracted off the auto-power estimates but it is zero for the cross-power spectra estimates between different redshift bins. $P_{b \ell}$ is a binning operator that bins the estimated $C_\ell$'s into bandpowers. For more details of the power spectrum estimator and its use, see \cite{brown05}. 

\section{Results} \label{sec:nu_results}

\begin{table*}
\begin{center}
\begin{tabular}{lccccc}
\hline
Experiment & Area [deg$^2$]&Frequency [MHz]& Resolution [arcsec]&FoM$_{\rm SNR}$&FoM$_{\rm DE}$\\
\hline
SKA1-Band 2-Centre 30\% & 5,000 & 1400&0.5&1652 &8.4 \\
SKA1-Band 1-Upper 30\% & 5,000 & 750&1.0&279 &1.2 \\
SKA1-Best Performing & 5,000 & 1050&0.55&2163 &10.2 \\
SKA2-Band 2-Centre 30\% & 30,000& 1400&0.5&1.4$\E^{5}$& 1.2$\E^{3}$ \\
\hline
\end{tabular}
\end{center}
\caption{Simulation results for some of the instrumental configurations considered, showing the relative performance of SKA1 Band 1, SKA1 Band 2 and SKA2.}\label{tab:results}
\end{table*}

\begin{figure*}
\includegraphics[width=\textwidth]{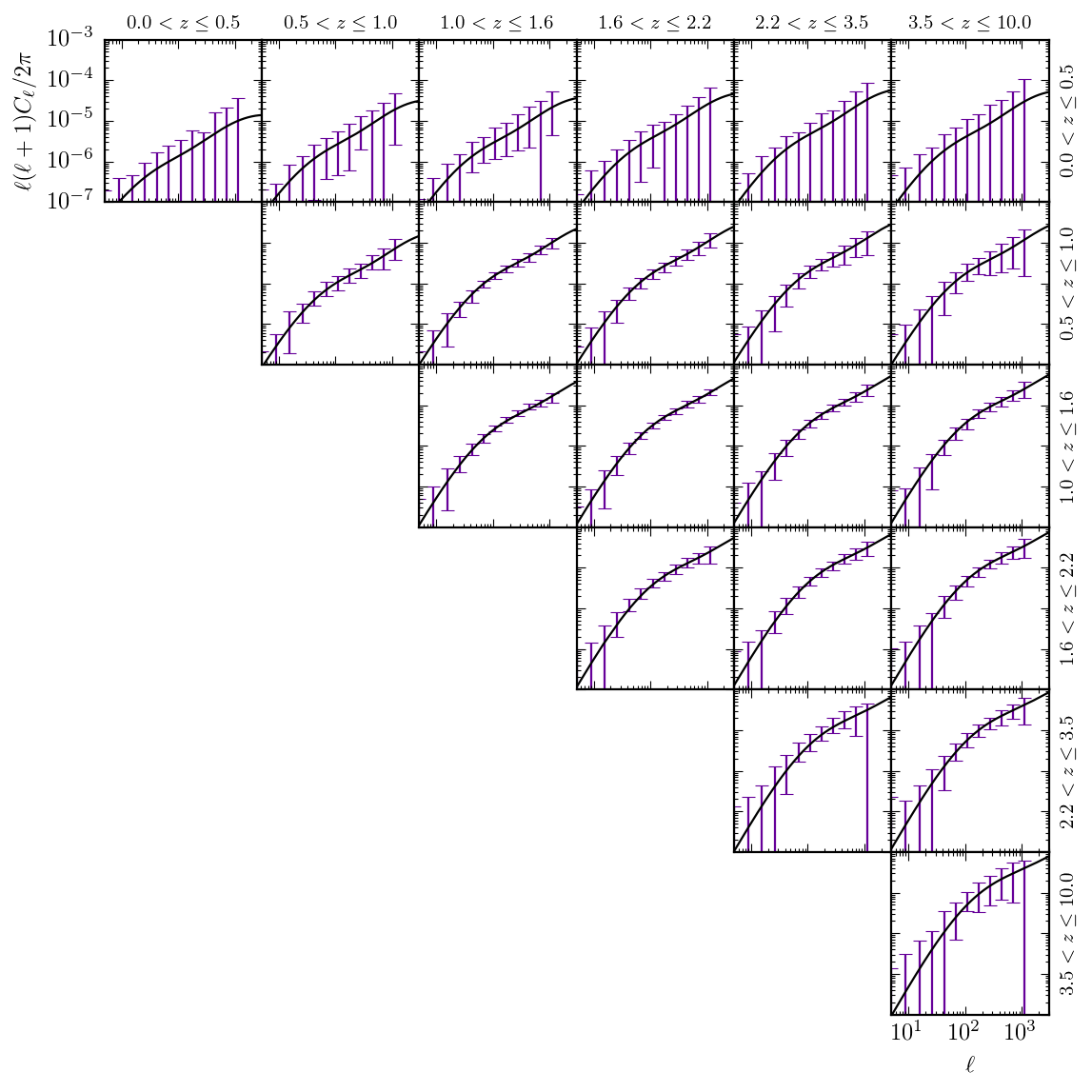}
\caption{Theory shear auto- and cross- power spectra for the six redshift bins (solid black lines) compared to recovered spectra averaged over 96 realisations (purple error bars) for the SKA1-Band 2-Centre 30\% experiment, with 0.5 arcsec resolution, a central frequency of $1400\, \mathrm{MHz}$ making use of 30\% of the available bandwidth.}
\label{fig:spectra}
\end{figure*}

\Cref{fig:spectra} presents the results of our simulation pipeline for one of the instrumental configurations and for the 6 redshift bins considered. The black lines are the input auto- and cross-spectra from theory and the purple error bars are the 1\,$\sigma$ confidence regions estimated from 96 realisation outputs. For each redshift slice, the power spectra are recovered for logarithmic bins in $\ell$, spaced by dlog$\ell=0.2$ between $\ell=1$ and $\ell=2048$.

For all the runs performed we obtained an unbiased recovery of the input shear power spectra. However, the size of the 1\,$\sigma$ confidence regions varies considerably with the experimental configuration. As a quantitative way to summarise this property, we defined a signal-to-noise figure of merit ${\rm FoM}_{\rm SNR}$ as:
\begin{equation}
{\rm FoM}_{\rm SNR}=\bmath{c}^{\rm T}\bmath{\sf Cov}^{-1}\bmath{c},
\label{eq:snrfom}
\end{equation}
where $\bmath{c}$ is a vector containing all the measured shear power spectra and cross-spectra for all bins in $\ell$ and redshift $\mathbf{c}=[\bmath{C}_{\ell}^{z1,z1},\,\bmath{C}_{\ell}^{z1,z2},...,\, \bmath{C}_{\ell}^{z6,z6}]$ and $\bmath{\sf Cov}$ is the covariance matrix of $\bmath{c}$ estimated from the simulation. We performed a large number $\sim50$ of pipeline runs in a grid covering the usable regions of the PSF size-frequency parameter space displayed in \cref{fig:rb_sens_plot} but here only present results for the best performing configurations (many configurations retained few enough galaxies after the resolution and sensitivity cuts such that the shear maps at our chosen resolution contained many holes, or no power spectrum detection was made). Finally, \cref{tab:results} summarises the results for some of the runs. 

\subsection{Optimising the observation frequency}
\begin{figure}
\includegraphics[width=0.5\textwidth]{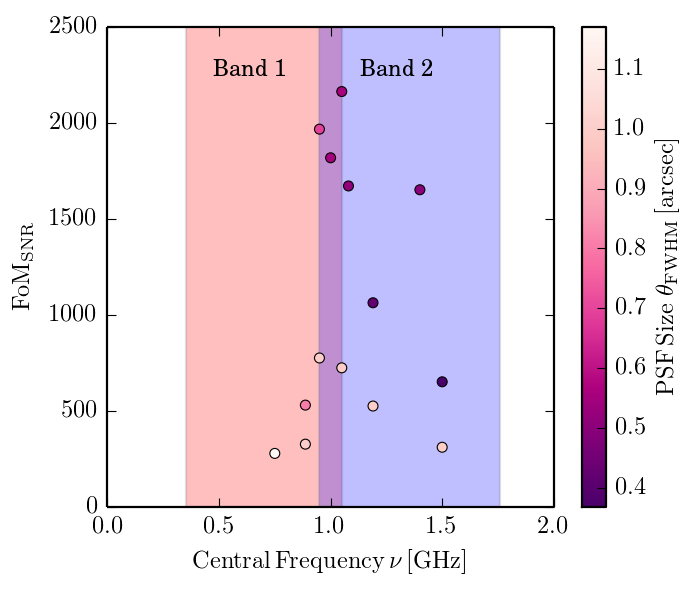}
\caption{Signal-to-noise figure of merit FoM$_{\rm SNR}$ shown for different frequencies (horizontal axis) and PSF sizes (colour scale) for the best-performing pipeline runs.} 
\label{fig:fom_freq}
\end{figure}

In \cref{fig:fom_freq} we show the figure of merit FoM$_{\rm SNR}$ as a function of frequency for the simulation runs performed for the baseline survey area of 5,000 deg$^2$. Our analysis clearly favours the higher frequency Band 2 with respect to Band 1. This is due to the lower resolution achievable in Band 1, which results in a significant reduction of the number of galaxies meeting the size cut. The higher source counts at lower frequencies are not able to compensate for this effect. 
We also show that best results are obtained when the central frequency of Band 2 is allowed to shift downwards to $\sim 1.1$\,GHz.  However, at these frequencies the improvement could  be  reduced  by  interference  with  the  Global  System for Mobile (GSM) band, which is not accounted for in our analysis.

\subsection{Use of full bandwidth}
We also consider the use of 100\% of the available SKA bandwidth, rather than only the central 30\%. This has the effect of increasing the sensitivity by a factor $\sqrt{3}$, but increases the influence of the ionosphere. In the case of Band 1, using the full bandwidth means the PSF size is now dominated by the ionospheric contribution at the lower limit of the band at $350 \, \mathrm{MHz}$, meaning few galaxies are resolved and several tomographic bins do not make a power spectrum detection. For Band 2, the figure of merit increases significantly as shown in \cref{fig:fom_area} (by the comparison between the red and fuchsia points). This increase should be weighed against the potential extra difficulty in modelling the shapes of sources with significant spectral indices across the band and potential loss of commensality with other science cases.

\subsection{Optimising the sky area}
\begin{figure}
\includegraphics[width=8cm]{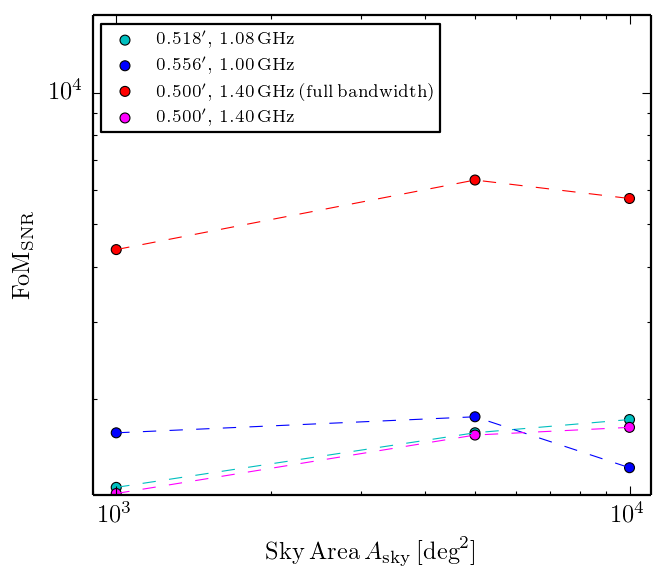}
\caption{Signal-to-noise figure of merit FoM$_{\rm SNR}$ as a function of sky areas for 4 configurations as detailed in the label. The comparison between the red points and the fuchsia points illustrate the improvement in the performance where the full Band 2 is used instead of the central 30\,\%.}
\label{fig:fom_area}
\end{figure}

The effect of different sky areas is shown in \Cref{fig:fom_area} for 4 instrumental configurations. 
The optimal survey area results to be 5,000 or 10,000 deg$^2$ depending on the run considered. However, we note that the curves showing FoM$_{\rm SNR}$ vs  survey area are quite flat, which suggests that the performance does not depend critically on the area of the survey. Specifically, for the runs considered, FoM$_{\rm SNR}$ changes by a factor 1.4--1.7 for an order of magnitude change in survey area.  { In fact, a reduction of the sky area obviously increases the size of the error bars at the lowest multipoles, which are poorly sampled. However, by spending more integration time per unit area, it also improves the sensitivity, thus reducing the error bars at high $\ell$'s. These two effects largely compensate each-other in FoM$_{\rm SNR}$.}

\subsection{Dark energy figure of merit}
\begin{figure}
\includegraphics[width=0.5\textwidth]{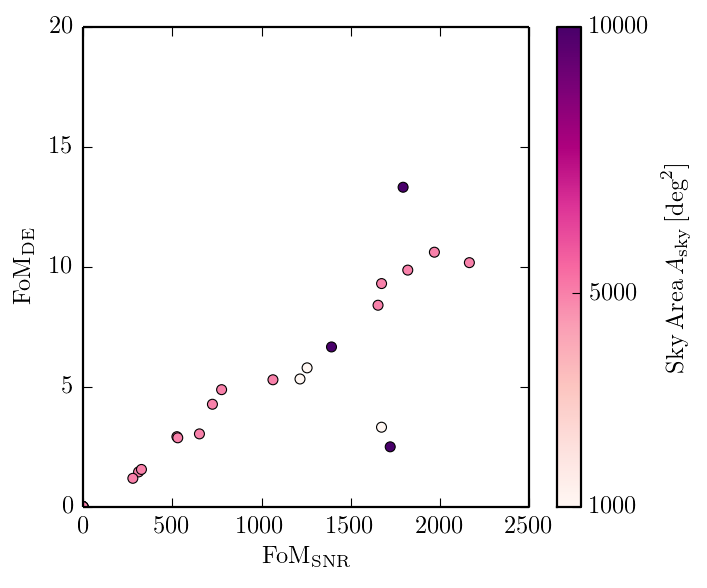}
\caption{Comparison between the signal-to-noise figure of merit \cref{eq:snrfom} and dark energy figure of merit \cref{eqn:defom}, showing a good correlation between the two.}
\label{fig:fom1vsfom2}
\end{figure}

The primary purpose of SKA weak lensing surveys will be to not just detect shear power spectra, but to measure cosmological parameters from them. Whilst the signal-to-noise figure of merit (Equation \ref{eq:snrfom}) provides a measure of how well we detect the shear power spectrum, this does not necessarily correlate well with the ability of a survey to do cosmological parameter estimation and model selection. In order to address this, we also calculate a dark energy figure of merit ${\rm FoM}_{\rm DE}$. As commonly defined in the literature, this quantity is proportional to the inverse of the area contained within the 1\,$\sigma$ marginal error contour on the dark energy equation-of-state parameters, $(w_0,w_a)$. To compute the dark energy FoM, we follow the Fisher matrix approach outlined in Paper I. In this case, though, the theoretical angular power spectra calculated according to \cref{eq:cl_theory} are calculated using the observed redshift distributions of sources obtained from the simulations. Then, we construct a Fisher matrix $\mathbfss F_{\alpha\beta}$ \citep[see e.g.][Equation 21]{harrisoncamera2015}, where greek indices label the dark energy parameter set $\vartheta_\alpha=\{w_0,w_a\}$. Then, the dark energy FoM is defined via
\begin{equation}
{\rm FoM}_{\rm DE}=\frac{1}{\sqrt{\det\left(\mathbfss F^{-1}\right)}}.
\label{eqn:defom}
\end{equation}
In \cref{fig:fom1vsfom2} we show the comparison between the signal-to-noise and the dark energy FoM. The two exhibit a very good correlation, which confirms that the signal-to-noise metric defined in \cref{eq:snrfom} can be used reliably to forecast cosmological results. As a consequence, the FoM$_{\rm DE}$ gives very similar indications as to what the optimal frequency and area of the survey is.

The dark energy FoMs we obtaine here for SKA1 (\cref{fig:fom1vsfom2} and \cref{tab:results}) are comparable with, though somewhat higher than, those reported in Paper I. This is because here we only vary $\lbrace w_0,w_a\rbrace$ and not the full set of standard $\Lambda$CDM parameters. This improvement is only partially compensated by the higher level of realism incorporated in this analysis, which reduces the number of galaxies and therefore increases errors. 

\subsection{SKA2 run}
\begin{figure*}
\includegraphics[width=\textwidth]{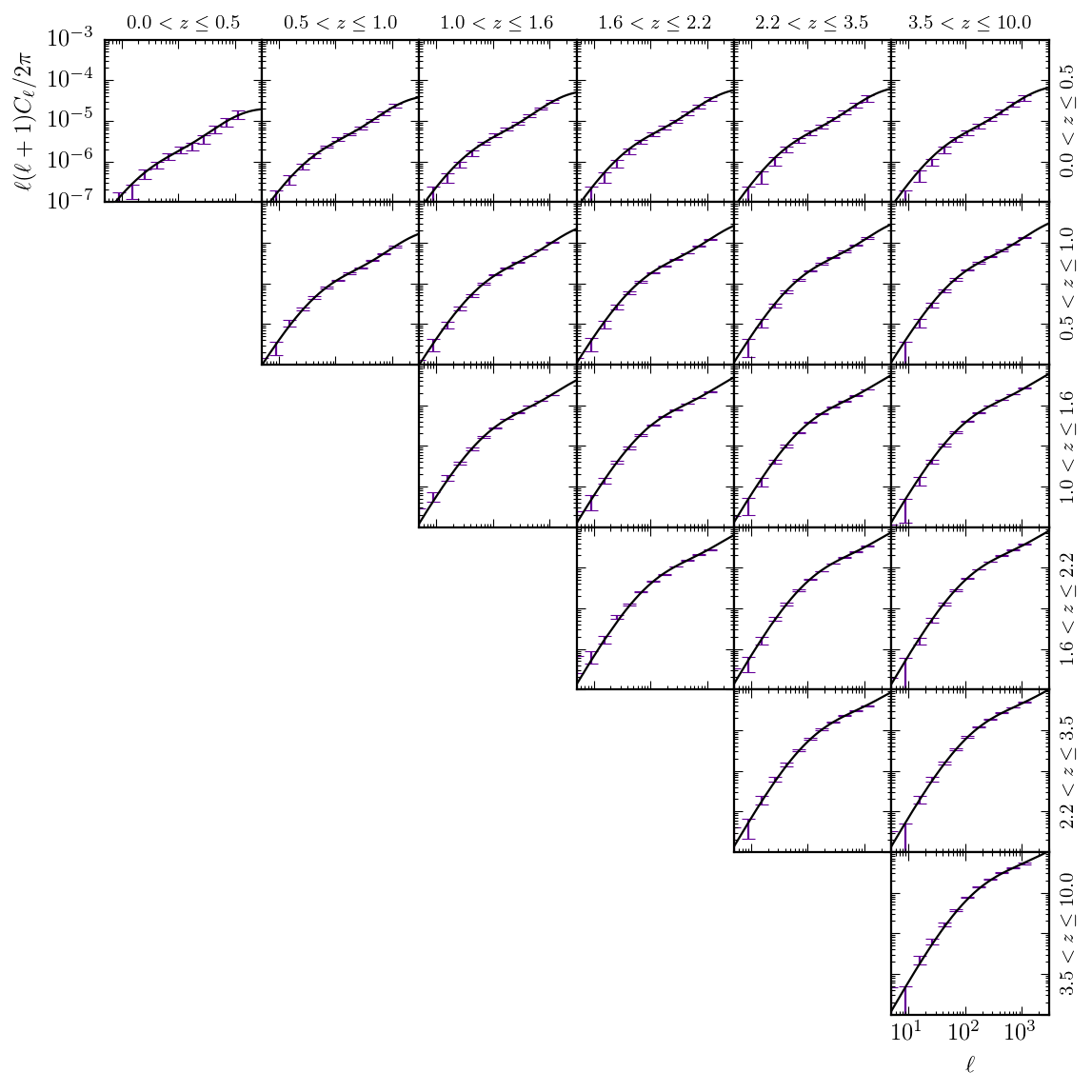}
\caption{The improvement on \cref{fig:spectra} from the factor ten improvement in sensitivity when considering the SKA2 telescope. Theory shear auto- and cross- power spectra for the six redshift bins (solid black lines) compared to recovered spectra averaged over 80 realisations (purple error bars) for the SKA2-Band 2-Centre 30\% experiment, with 0.5 arcsec resolution, a central frequency of $1400\, \mathrm{MHz}$ making use of 30\% of the available bandwidth.}
\label{fig:spectra_ska2}
\end{figure*}

In \cref{fig:spectra_ska2} we show the spectra and cross-spectra for the 6 redshift bins for the second phase deployment of the SKA (SKA2-MID). { In this case, we consider just one instrumental configuration, as a demonstration of what can be achieved with the full SKA. Consistently with what considered in \cite{2015arXiv150103828B}}, this run corresponds to a 30,000 deg$^2$ sky coverage and a 10,000 hours integration time, with central frequency of 1400\,MHz and half-arcsecond resolution. { We assume to have spectroscopic redshifts for 50\,\% or the sources below redshift 2.0, and errors on the photometric redshifts as before.}

The recovery of the spectra is excellent and yields figures of merit of FoM$_{\rm SNR}=1.4\,\times 10^{5}$ and FoM$_{\rm DE}=1.2\,\times 10^{3}$. Note that the dark energy FoM should be considered as an upper limit, because the total error for such a precise measurement is dominated by degeneracies with other standard cosmological parameters, which are not accounted for in the present analysis. Nonetheless, our result is broadly consistent with what found in our companion work, where Section 4.2 shows the case for a forecast where no marginalisation over the full set of standard $\Lambda$CDM parameters has been applied (see values quoted in the caption of Table 2, Paper I). After such a marginalisation, the dark energy figure of merit drops to values $\sim$50.

\section{Conclusions}
\label{sec:conclusions}
We have developed a detailed simulation pipeline to assess the capabilities of the SKA to perform accurate weak lensing studies. Our simulations takes into account several real world effects:
\begin{enumerate}
\item realistic radio-source properties (correlated redshift, flux and size distributions);
\item a selection function based on realistic sensitivity and resolution cuts implemented for different SKA configurations;
\item the effect of the ionosphere;
\item errors in the galaxy shape measurement dependent on the source signal-to-noise; 
\item errors in the measurement of redshift for the sources in the catalogue;
\item a realistic power spectrum estimation procedure in the presence of incomplete sky coverage.
\end{enumerate}
Such a detailed simulation goes significantly beyond previous weak lensing forecasting studies. We applied our simulation tool to optimise a weak lensing survey for the first deployment phase of the SKA (SKA1-MID). We based our assessment on the total signal-to-noise of the shear power spectra, a metric that we found to correlate very well with a standard dark energy figure of merit.

With a 5,000 deg$^2$ survey, best results are obtained with Band 2 (covering 950-1760\,MHz as currently defined) with respect to Band 1 (covering 350-1050\,MHz). This is mostly due to the higher resolution achievable in Band 2, which is not compensated by the higher source counts achievable in Band 1. Indeed, the best trade-off between sensitivity, resolution and source counts would be obtained by shifting the centre of Band 2 to lower frequencies, $\sim$1100\,MHz. This however does not take into account GSM interference, which could reduce the improvement.

In terms of survey area, best results are obtained with 5,000 deg$^2$ or 10,000 deg$^2$, depending on the instrumental configuration. The effect of survey area is however not as important as that of the central frequency of the survey. Finally, we also presented the forecasts for the second deployment phase of the SKA (SKA2-MID) for a 30,000 deg$^2$ survey in Band 2 as currently defined. This improves the dark energy figure of merit of two orders of magnitude with respect to the same instrumental configuration for SKA1-MID. 

\section*{Acknowledgments}

The authors are supported by an ERC Starting Grant (grant no. 280127). MLB is an STFC Advanced/Halliday fellow. We wish to thank Tomasz Kacprzak for making the \imshape\ simulations available to us and Wendy Williams and Huub R\"ottgering for useful discussions on the ionosphere. We also thank Robert Braun for help with the SKA sensitivity curves.

\bibliography{ska_wl_sims}
\bibliographystyle{mn2e_plus_arxiv}

\bsp

\label{lastpage}

\end{document}